\documentclass[aps,pre,twocolumn,amsmath,amssymb,superscriptaddress]{revtex4-2}
\usepackage[colorlinks]{hyperref}
\usepackage{graphicx}
\usepackage[dvipsnames]{xcolor}
\usepackage{braket}
\usepackage{soul}
\usepackage{bbm}
\usepackage{url} 
\usepackage{ulem}
\usepackage{amsmath}
\usepackage{algorithm}
\usepackage{algpseudocode} 

\definecolor{blue1}{rgb}{0.1294 0.1961 0.8431}
\definecolor{green1}{rgb}{0.16, 0.38, 0.27}
\definecolor{red1}{rgb}{0.8500 0.3250 0.0980}
\definecolor{purple1}{rgb}{0.4940 0.1840 0.5560}
\definecolor{black1}{rgb}{0 0 0}

\definecolor{green2}{rgb}{0.7020    0.9137    0.6784}
\definecolor{green3}{rgb}{0.2275    0.7333    0.2667}
\definecolor{green4}{rgb}{0.1600    0.3800    0.2700}

\definecolor{blue1}{rgb}{ 0    0.7843    1}
\definecolor{blue2}{rgb}{0    0.3922    1}
\definecolor{blue3}{rgb}{0     0     1}

\definecolor{red1}{rgb}{1   0.7843         0}
\definecolor{red2}{rgb}{1    0.3922         0}
\definecolor{red3}{rgb}{1   0      0}

\definecolor{fig2blue}{rgb}{0 0.4470 0.7410}
\definecolor{fig2red1}{rgb}{1.0000    0.9412    0.1098}
\definecolor{fig2red2}{rgb}{1.0000    0.5451    0.1373}
\definecolor{fig2red3}{rgb}{1.0000    0.0941    0.1569}

\newcommand{\sutdphys}{Science, Mathematics and Technology Cluster, Singapore
University of Technology and Design, 8 Somapah Road, 487372 Singapore}
\newcommand{\sutdepd}{EPD Pillar, Singapore University of Technology and Design, 8 Somapah Road, 487372 Singapore}

\newcommand{\cqt}{Centre for Quantum Technologies, National University of Singapore 117543, Singapore} 
\newcommand{\majulab}{MajuLab, CNRS-UNS-NUS-NTU International Joint Research Unit, UMI 3654, Singapore}

\begin{document}
	
\title{Concentrated Monte Carlo sampling for local observables in quantum spin chains}    
 
\author{Wenxuan Zhang}  
\affiliation{\sutdphys}

\author{Dingzu Wang}  
\affiliation{\sutdphys}

\author{Dario Poletti} 
\affiliation{\sutdphys}
\affiliation{\sutdepd}
\affiliation{\cqt}
\affiliation{\majulab}

\begin{abstract}  

Monte Carlo methods are widely used to estimate observables in many-body quantum systems. However, conventional sampling schemes often require a large number of samples to achieve sufficient accuracy. 
In this work we propose the concentrated Monte Carlo sampling approach, which builds on the idea that in systems with only short range correlations, to obtain accurate expectation values for local observables, one would favor detailed information in the surroundings of this observable compared to far away from it. 
In this approach we consider all possible configurations in the surroundings of a local observable, and unique samples from the remaining of the setup drawn using Markov chain Monte Carlo. 
We have tested the performance of this approach for ground states of the spin-1/2 tilted Ising model in different phases, and also for thermal states in the a spin-1 bilinear-biquadratic model.    
Our results demonstrate that CMCS yields higher accuracy for local observables in short-range correlated states while requiring substantially fewer samples, showcasing in which regimes one can obtain acceleration for the evaluation of expectation values.  
\end{abstract}
\date{\today}
\maketitle

\section{Introduction} 
Monte Carlo(MC) simulations have become a standard tool for estimating observables in statistical physics by sampling configurations from a target measure. 
Its reach spans statistical mechanics \cite{binder1997applications, newman1999monte, landau2021guide}, molecular electronic-structure theory \cite{ceperley1980ground, foulkes2001quantum, austin2012quantum}, quantum lattice systems \cite{hirsch1985two, sandvik1991quantum, sandvik2010computational, capogrosso2008monte, sandvik2007evidence, varney2009quantum, mazurenko2017cold}, and neural quantum states \cite{carleo2017solving, carleo2019machine, hibat2020recurrent, sharir2020deep, zhang2025paths, gravina2025neural, van2025many, sinibaldi2024timedependentneuralgalerkinmethod, WangPoletti2025}; 
and has also led to the development of specialized MC algorithms that improve sampling efficiency and accuracy for classical and quantum many-body systems, such as Markov chain Monte Carlo (MCMC) \cite{metropolis1953equation, hastings1970monte}, 
stochastic series expansion algorithm \cite{sandvik1999stochastic, syljuaasen2002quantum}, worm algorithm \cite{prokof2001worm, prokof1998worm}, auxiliary-field quantum Monte Carlo \cite{blankenbecler1981monte, zhang1997constrained, zhang2003quantum}, path-integral Monte Carlo \cite{feynman2018statistical, ceperley1995path, boninsegni2006worm}, variational Monte Carlo \cite{foulkes2001quantum, becca2017quantum, toulouse2016introduction}. With MC approaches, one can estimate large sums or integrals by sample averages over configurations drawn from a target distribution, and the standard error decreases as $1/\sqrt{N_s}$, where $N_s$ is the number of samples, independent of the system’s dimension. 

\begin{figure}
	\centering
	\includegraphics[width=0.95\linewidth]{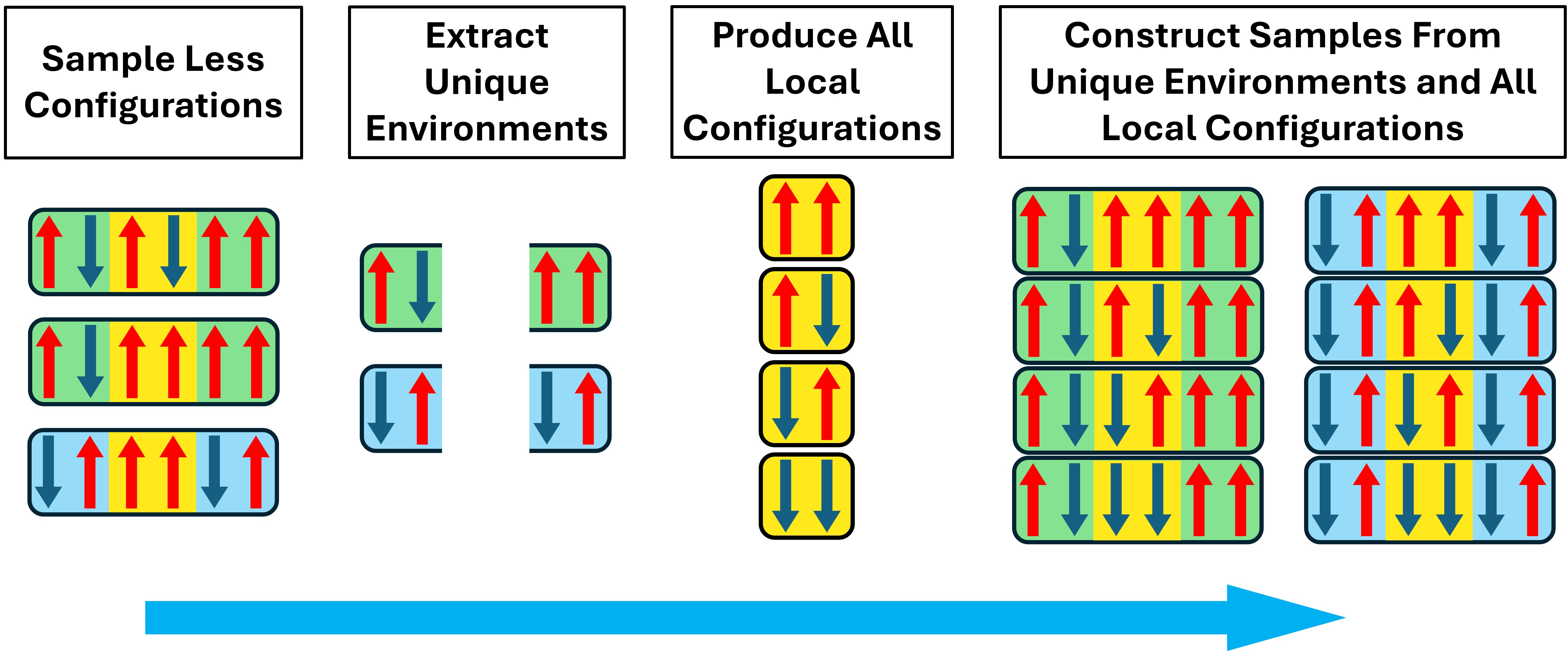}
	\caption{Illustration of CMCS algorithm for a one-dimensional spin-1/2 chain with 6 sites. 
    Red (dark blue) arrows denote spin-up (spin-down) states. 
    Each Monte Carlo sample is partitioned into a local region, in this case the 2 central sites, and the environment, in this case consisting of the remaining 4 sites, depicted with different background colors according to the distinct environment configurations. 
    For the local region, all possible configurations are produced, and they are recombined with the unique environment configurations to form the configurations from the concentrated sampling. 
    } 
 \label{fig1}
\end{figure}

In many physical applications, the observables of interest are local, and the underlying systems often exhibit only short-range correlations. 
For these systems, a deeper knowledge of the local properties of the system could lead to more accurate estimators, despite reducing the amount of information away from the local region. 
Motivated by this, we introduce {\it Concentrated Monte Carlo Sampling} (CMCS) which is depicted in Fig. \ref{fig1}. CMCS partitions each global sample into a local region and an environment, collects the environments into a set of distinct configurations, and then pairs each with all configurations of the chosen local region to form a set of unique configurations. These reconstructed configurations are assigned appropriate weights and used to evaluate local observables. 
To explore the performance of this method, we first consider the ground state of the one-dimensional non-integrable tilted Ising model (TIM). We analyze the performance of CMCS for local observables such as short-range spin correlations across the antiferromagnetic (AFM), ferromagnetic (FM), and paramagnetic (PM) phases, where the correlation length differs markedly. We then consider finite-temperature states of a one-dimensional spin-1 bilinear–biquadratic chain, which exhibits, in its ground state, a rich phase diagram due to its larger on-site Hilbert space and competing interactions, and study how the accuracy of the expectation values from CMCS depends on temperature. 
We find that indeed for system with shorter-range correlations, the CMCS approach leads to higher accuracy with less total number of samples, as long as we compute observable within the region where we concentrate our sample.
This is true also in ferro and anti-ferromagnetic phases, while in the critical region the concentrated sampling approach cannot match traditional methods. 
For finite-temperature systems, the performance of CMCS are significantly better at higher temperatures and can remain competitive for regions of lower temperatures.  
In Sec.~\ref{sec:cmcs} we describe the method in detail, in Sec.~\ref{sec:groundstate} we detail its performance for ground states, while in Sec.~\ref{sec:thermalstate} we study thermal states and, finally, in Sec.~\ref{sec:conclusions} we draw our conclusions. 

\section{Concentrated Monte Carlo Sampling} \label{sec:cmcs}
In this work, we consider a one-dimensional quantum spin chain of length $L$. A physical configuration is denoted by
\begin{equation}
    \pmb{x} = \left[x_1, x_2, ..., x_L\right],
\end{equation}
where $x_i$ represents the local state at site $i$. 
To evaluate observables, the probability of each configuration is required. As the Hilbert space grows exponentially with system size, it is impossible to enumerate all configurations explicitly. Instead, the system is represented by a probability distribution $P(\pmb{x})$, from which configurations can be sampled.
Here, we choose MCMC as our sampling scheme. In particular, we first generate a set of samples $\mathcal{S} =\{\pmb{x}^{(1)}, \pmb{x}^{(2)}, ...,\pmb{x}^{(N_s)}\}$ according to the target distribution $P$, where $N_s$ is the total number of samples. 

Since we are interested in local observables, which act only on a restricted subset of sites, we refer to these sites as the local sites. These sites are embedded into a {\it local region}, which always includes all local sites and may also contain additional neighboring sites. All the remaining sites outside of the local region are referred to as the {\it environment}. Hence, each sampled global configuration $\pmb{x}$ is decomposed as
\begin{equation}
    \pmb{x} = (\pmb{x}_{loc}, \pmb{x}_{env})
\end{equation}
where $\pmb{x}_{loc}$ is the configuration for the local region and $\pmb{x}_{env}$ is the configuration for the environment.

Within the local region, we will then enumerate all possible configurations $\{\pmb{u}_{loc}^{(i)}\}_{i=1}^{N_{\ell}}$, which span a complete subspace of dimension $N_{\ell} = d^{\ell}$, where $d$ represents the Hilbert space dimension per site and $\ell\in[1,L]$ is the size of the local region. From the sample set $\mathcal{S}$, we extract the environment configurations $\{ \pmb{x}^{(k)}_{env}\}_{k=1}^{N_s}$ and remove duplicates, resulting in a set of unique environment configurations $\{ \pmb{u}_{env}^{(j)} \}_{j=1}^{N_e}$, where $N_e$ is the number of unique environment configurations. Each  $\pmb{u}_{env}^{(j)}$ is then combined with $\{\pmb{u}_{loc}\}$, thereby constructing a new ensemble of $N_{u} = N_{\ell} \times N_e$ unique configurations 
$\mathcal{U}$, where each configuration is $\pmb{u}^{(i,j)} = (\pmb{u}_{loc}^{(i)}, \pmb{u}_{env}^{(j)})$, and $N_u$ is the total number of unique configurations. 
This procedure is depicted in Fig.~\ref{fig1}. 

Given the ensemble $\mathcal{U} = \{ \pmb{u}^{(i,j)} \}$ of unique configurations constructed above, the probability distribution has to be renormalized within this subset, with the weight of each $\pmb{u}$ defined as the renormalized form of its original value,
\begin{equation}
    \tilde{P}(\pmb{u}) = \frac{P(\pmb{u})}{\sum_{\pmb{y} \in \mathcal{U}}P(\pmb{y})}. 
\end{equation}

The expectation value of the local observable $O_{loc}$ is then estimated as 
\begin{equation}
    \langle O_{loc} \rangle = \sum_{\{\pmb{u}\}}\tilde{P}(\pmb{u})O_{loc}(\pmb{u})
\end{equation}

The main steps of this method are described by the pseudo-code in Alg.~\ref{alg:CMCS}.  
In the limit of large $N_s$ and large $N_e$, $\tilde{P}$ converges towards the probability $P$, thus providing an estimator for it. What we investigate in the next sections is in which conditions one can obtain accurate local observables $\langle O_{loc} \rangle$ with the concentrated sampling compared to using the samples from MCMC.

\begin{algorithm}[H]
\caption{Concentrated Monte Carlo Sampling}
\label{alg:CMCS}
\begin{algorithmic}[1]

    \State \textbf{Sampling via Monte Carlo:} Generate Monte Carlo sample set $\mathcal{S}=\{\pmb{x}^{(k)}\}_{k=1}^{N_s}$ from probability distribution $P(\pmb{x})$.
    
    \State \textbf{Partition:} split every sample as the local part $\pmb{x}_{loc}$ and the envionment part $\pmb{x}_{env}$: $\pmb{x}^{(k)} = (\pmb{x}^{(k)}_{loc}, \pmb{x}^{(k)}_{env})$

        \Statex \hspace{\algorithmicindent}\parbox[t]{0.85\linewidth}{2.1: \textbf{Enumerate local subspace:} Define a local region and enumerate all possible local configurations: $\{\pmb{x}_{loc}^{(k)}\}_{k=1}^{N_s} \rightarrow \{\pmb{u}_{loc}^{(i)}\}_{i=1}^{N_{\ell}}$.}

        \Statex \hspace{\algorithmicindent}\parbox[t]{0.85\linewidth}{2.2: \textbf{Unique environments:} Collect samples $\pmb{x}_{env}$ and extract the unique environment configurations: $\{\pmb{x}_{env}^{(k)}\}_{k=1}^{N_s} \rightarrow 
        \{\pmb{u}_{env}^{(j)}\}_{j=1}^{N_e}$.} 
        
    \State \textbf{Reconstruction:} Construct the set of unique global (local plus environment) samples $\{\pmb{u}^{i,j}\}_{i,j=1}^{N_{\ell},N_e}$.

    \State \textbf{Renormalization}: calibrate the probability of each $\pmb{u}$ evaluating $\tilde{P}(\pmb{u})$.
    
    \State \textbf{Estimate observable:} Return $\langle O_{loc} \rangle = \sum\tilde{P}(\pmb{u})O_{loc}(\pmb{u})$

\end{algorithmic}
\end{algorithm}

\section{Ground State of Tilted Ising Model} \label{sec:groundstate}
To investigate the performance of CMCS, we study both ground state and thermal states. In this section, we first evaluate the local observables in the ground states of a one-dimensional tilted Ising model (TIM) with open boundary condition. The Hamiltonian is given by
\begin{align}
    \mathcal{H} &= J\sum_{i}\sigma^z_i\sigma^z_{i+1} - \sum_i(h_x \sigma^x_i + h_z\sigma^z_i)\label{eq:tim} 
\end{align}
where, $\sigma_i^a(a = x,y,z)$ are Pauli operators of $i$-th site; $J$ is the nearest-neighbor coupling strength, and $h_x,\; h_z$ represent the external magnetic fields along $x$- and $z$-directions, respectively. When the local fields are small in magnitude, the ground state of the system is either in the ferromagnetic phase when $J<0$ or in the antiferromagnetic phase when $J>0$; when local fields are large enough compared to $J$, the ground state of the system is in the paramagnetic phase.

Since the purpose is to evaluate the effectiveness of the sampling, we evaluate very accurately the ground state and its obervables using the matrix product state (MPS) method \cite{white1992density, schollwock2011density} to produce both a reliable wave-function to sample from, and the ground truth for value for the observables. 
We will focus on the single-site observable $\langle \sigma^z_i\rangle$, which represents the expectation value of local magnetization in the $z$-direction, as well as two-site observables $\langle \sigma^z_i \sigma^z_{i+1} \rangle$, corresponding to the spin-spin correlation in the $z$-direction.
For each quantity, we perform $R=10$ independent runs and compute the mean absolute error $ \epsilon (\langle O_{loc}\rangle) = \frac{1}{R}\sum_{r}|\langle O_{loc} \rangle - \langle O_{loc} \rangle_{\text{ex}}|$, where $\langle O_{loc} \rangle_{\text{ex}}$ is the exact value, here from MPS. Specifically, we focus on a 20-spin TIM with $2^{20}$ configurations, which provides a sufficiently large Hilbert space such that the sampling remains nontrivial.

\begin{figure}
	\centering
    \includegraphics[width=\linewidth]{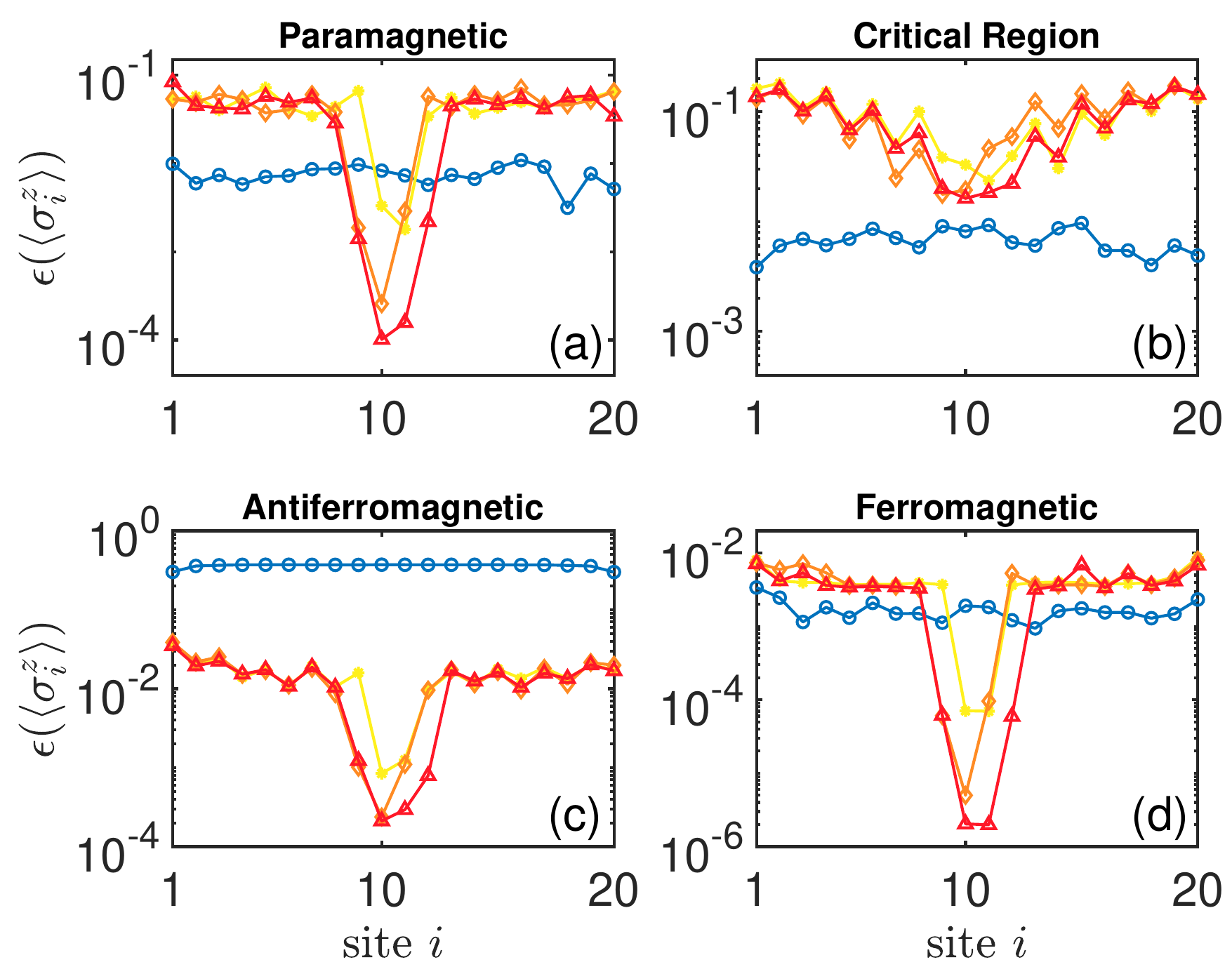}
	\caption{Performance of CMCS and MCMC for the different ground states of 1D 20 spins tilted Ising model, presented for different local region sizes $\ell$. The panels show the mean absolute error $\epsilon$ of $\langle \sigma^z_i \rangle$ as a function of the site index $i$, where the exact values are obtained by MPS. The system is in 
    (a) the paramagnetic phase with $h_x=10J,\; h_z=J/2$, 
    (b) the critical region with $h_x=0.95J, \;h_z=J/2$, 
    (c) the antiferromagnetic phase with $h_x=J/2, h_z=J/2$ and 
    (d) the ferromagnetic phase with $J<0, h_x=|J|/2,\; h_z=|J|/2$.
    \textcolor{fig2blue}{$\circ$} represents MC sampling with $N_s=10^4$;
    \textcolor{fig2red1}{$\ast$} represents CMCS $\ell = 2$ with sites 10$-$11, $N_l=4$ and $100\le{N_{u}}\le 2000$;
    \textcolor{fig2red2}{$\lozenge$} represents CMCS $\ell = 3$ with sites 9$-$11, $N_l=8$ and $ 192\le{N_{u}} \le 4000$;
    \textcolor{fig2red3}{$\bigtriangleup$} represents CMCS $\ell = 4$ with sites 9$-$12, $N_l=16$ and $  352\le{N_{u}} \le 8000$.
    }
 \label{fig2}
\end{figure}

\begin{figure}
	\centering
    \includegraphics[width=\linewidth]{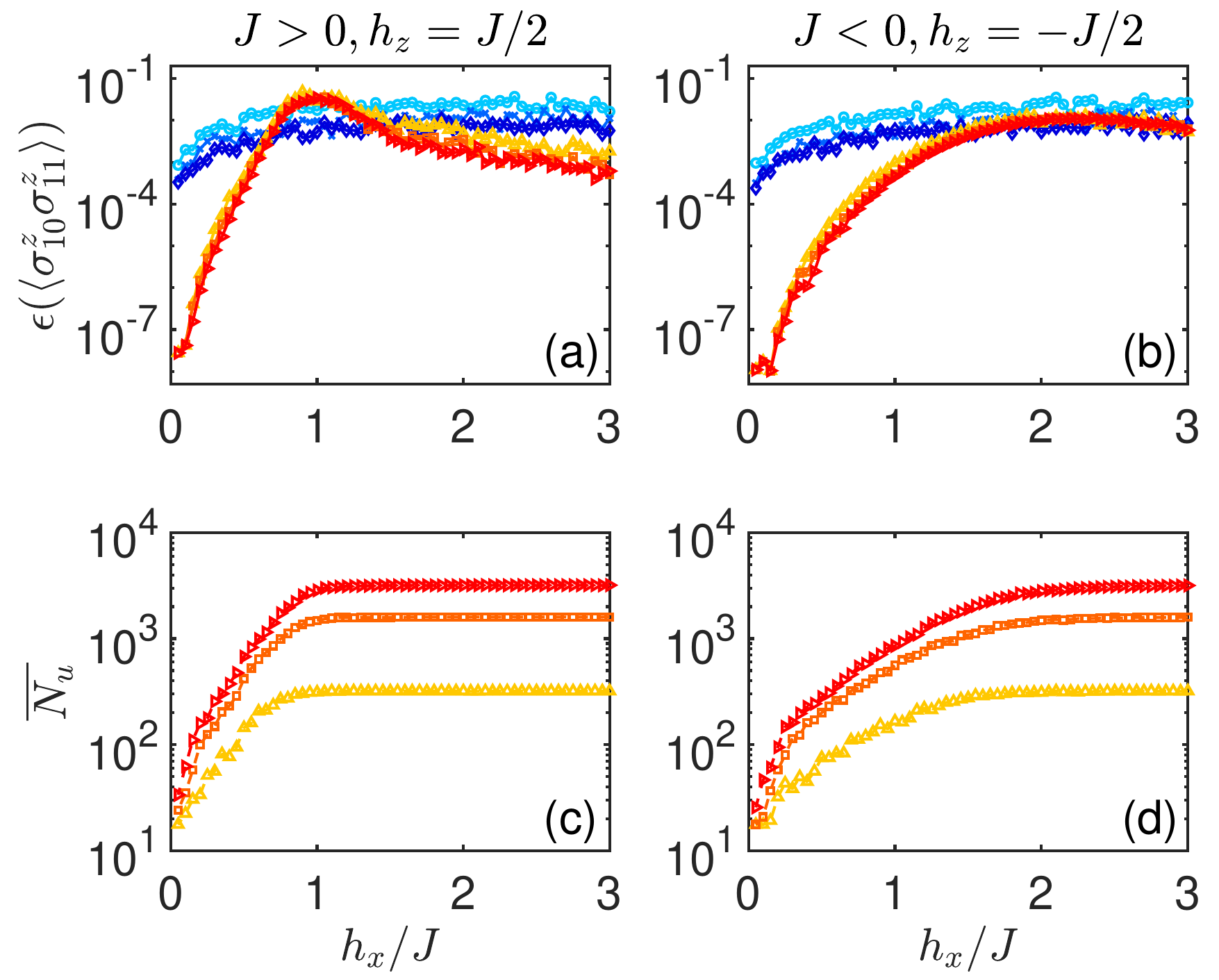}
	\caption{Comparison between CMCS and MCMC for the ground state of 1D 20 spins tilted Ising model, at different values of $h_z$ and for $J>0$, panels (a) and (c), and $J<0$, panels (b) and (d), while we consider a fixed local region size $\ell=4$ and fix $h_z=|J|/2$. Results are shown for varying sample sizes. The top panels show the mean absolute error $\epsilon$ of the correlation between the two central sites $\langle \sigma^z_{10} \sigma^z_{11}\rangle$, where exact values are obtained by MPS. The bottom panels show the average number of unique configurations, $\overline{N_u}$. 
    Blue markers indicate the different samples in MC: 
    \textcolor{blue1}{$\circ$} represents $N_s = 1000$; 
    \textcolor{blue2}{$\times$} represents $N_s = 5000$; 
    \textcolor{blue3}{$\lozenge$} represents $N_s = 10000$. 
    Red markers indicates the different samples in CMCS: 
    \textcolor{red1}{$\triangle$} represents $N_s = 20$; 
    \textcolor{red2}{$\square$} represents $N_s = 100$; 
    \textcolor{red3}{$\triangleright$} represents $N_s = 200$. (Here, $N_s$ in CMCS denotes the number of raw MCMC samples used to construct the unique environments, not the number of unique configurations.) 
    } 
 \label{fig3}
\end{figure}

\begin{figure*}
	\centering
    \includegraphics[width=\linewidth]{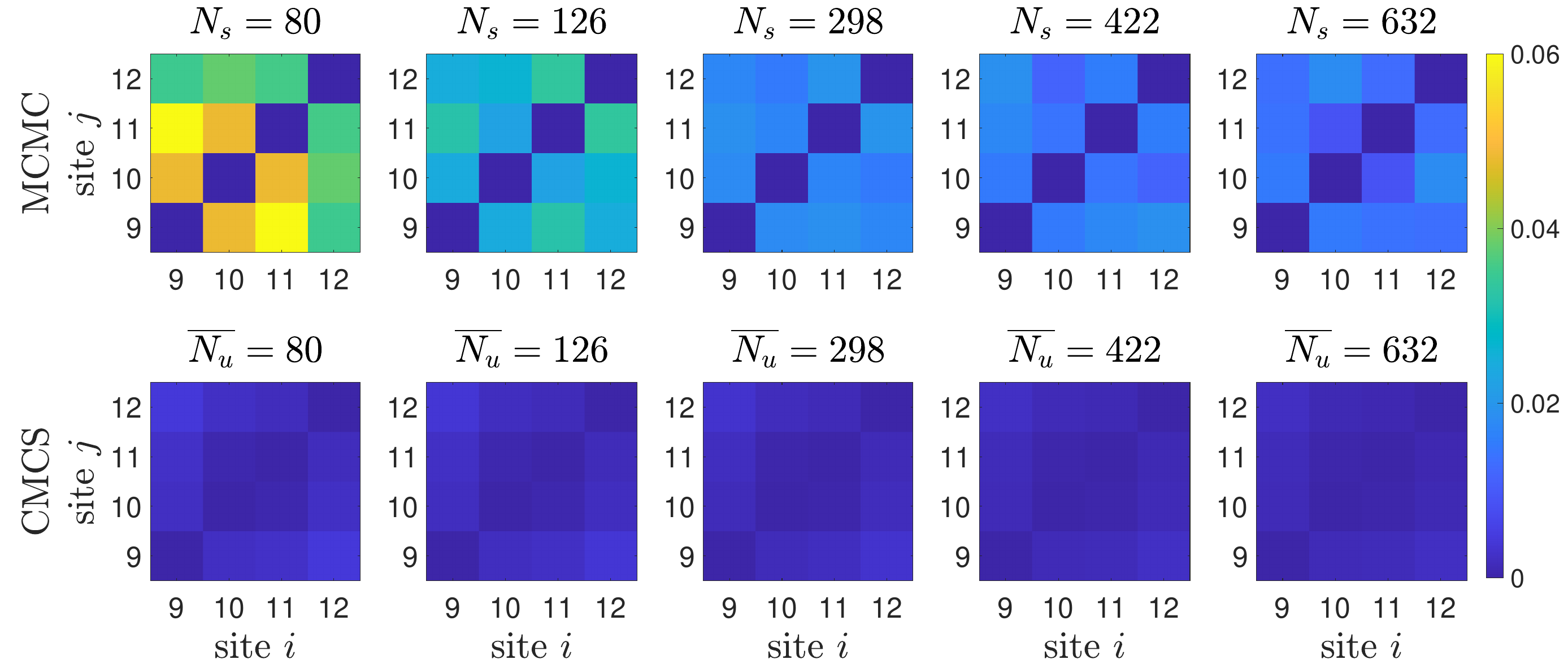}
	\caption{Comparison of the mean absolute error of correlation matrix $\langle \sigma^z_i \sigma^z_j \rangle$ between MCMC and CMCS for the ground state of 1D 20 spins tilted Ising model for the local region size $\ell = 4$ with sites 9$-$12. Results are shown for varying sample sizes $N_s$ and average numbers of unique configurations $\overline{N_u}$. The system is in the antiferromagnetic phase, with parameters $h_x=J/2, h_z=J/2$. The top panels present the mean absolute error $\epsilon$ of the correlation matrix of the local region obtained from MCMC, while the bottom panels show the corresponding results from CMCS. } 
 \label{fig4}
\end{figure*} 

In Fig.~\ref{fig2}, we benchmark the performance of CMCS against conventional MCMC across different ground states by plotting the mean absolute error of $\langle \sigma^z_i \rangle$ as a function of the site index $i$. 
In all panels, the local region used in CMCS is placed in the center of the chain and its size $\ell$ is varied as indicated.
We can see that, except near criticality, CMCS outperforms MCMC within the chosen central local region, whereby local observables evaluated inside the local region exhibit significantly smaller errors. 
Specifically, in panel (a), we consider the system to be in the paramagnetic phase, where the probability distribution in the $\sigma^z$ basis is close to uniform and the true value of $\langle \sigma^z_i\rangle$ is very small but not exactly 0. 
In this case, MCMC suffers from statistical noise of order $1/\sqrt{N_s}$ that can overwhelm the signal, so the small but finite expectation value cannot be statistically resolved from zero at the current sample size.  
In contrast, the same near-uniform distribution means that most environment configurations have very similar weights, and from the perspective of the CMCS algorithm, the local observable is therefore weakly sensitive to the specific environment configurations. This increases the effective sample size and suppresses the noise, so that these near-zero values of local observable are accurately resolved by CMCS. 
In the ordered phases shown in panels (c) and (d), the probability weight is concentrated around a few ordered configurations. 
For local-update MCMC this leads to long autocorrelation times and the Markov chain is easily trapped near local minima. 
Obtaining reliable results with MCMC in this case would therefore require much longer chains and many more samples, although it can be improved with importance sampling techniques \cite{alain2016variancereductionsgddistributed,
katharopoulos2019samplescreatedequaldeep, misery2025lookingelsewhereimprovingvariational, yang2020deep}. 
CMCS is well suited to such highly concentrated distributions, as it still explores the configurations within a local region,  
yielding precise local observables even when the initial number of Monte Carlo samples is relatively small.        
Around the critical region shown in panel (b), the correlation length becomes large and local observables become sensitive to distant parts of the system. Insufficient sampling of the environment configurations can then have a strong impact on the local observables. Under these conditions, CMCS generally does not outperform conventional MCMC. Similar results are obtained when measuring $\langle \sigma^x_i\rangle$, see App.~\ref{app}.

In Fig.~\ref{fig3}, we compare the performance of CMCS and MCMC for the correlation in the two central sites $\langle\sigma^z_{10}\sigma^z_{11}\rangle$ across varying $h_x$ and sample sizes. 
In top two panels we display the error, while in the bottom two we show the number of unique configurations used by the CMCS approach $\overline{N_u}$, averaged over 10 realizations. We can see that even with a large sample size of $N_s=10^4$, MCMC errors still remains around $10^{-3}$, although it is not strongly dependent on $h_x$, whereas CMCS achieves comparable or smaller errors with far fewer effective samples away from criticality.

In Fig.~\ref{fig4}, we compare MCMC and CMCS with the same sample size, though evaluating the correlation matrix of the entire local region for a system in the ferromagnetic phase; the results are displayed as a $4 \times 4$ panel where each cell reports a specific pair $(i,j)$. 
We can see that using the same number of samples, CMCS consistently attains lower errors than MCMC. 
Beyond reducing overall error, CMCS also makes the error distribution across the local correlation matrix substantially more uniform, whereby errors for each element of the correlation matrix are significantly reduced, and the disappearance of large off-diagonal errors. Consequently, CMCS can improve not only a single nearest neighbor but the entire set of two-point correlators within the local region.

\begin{figure}
	\centering
	\includegraphics[width=\linewidth]{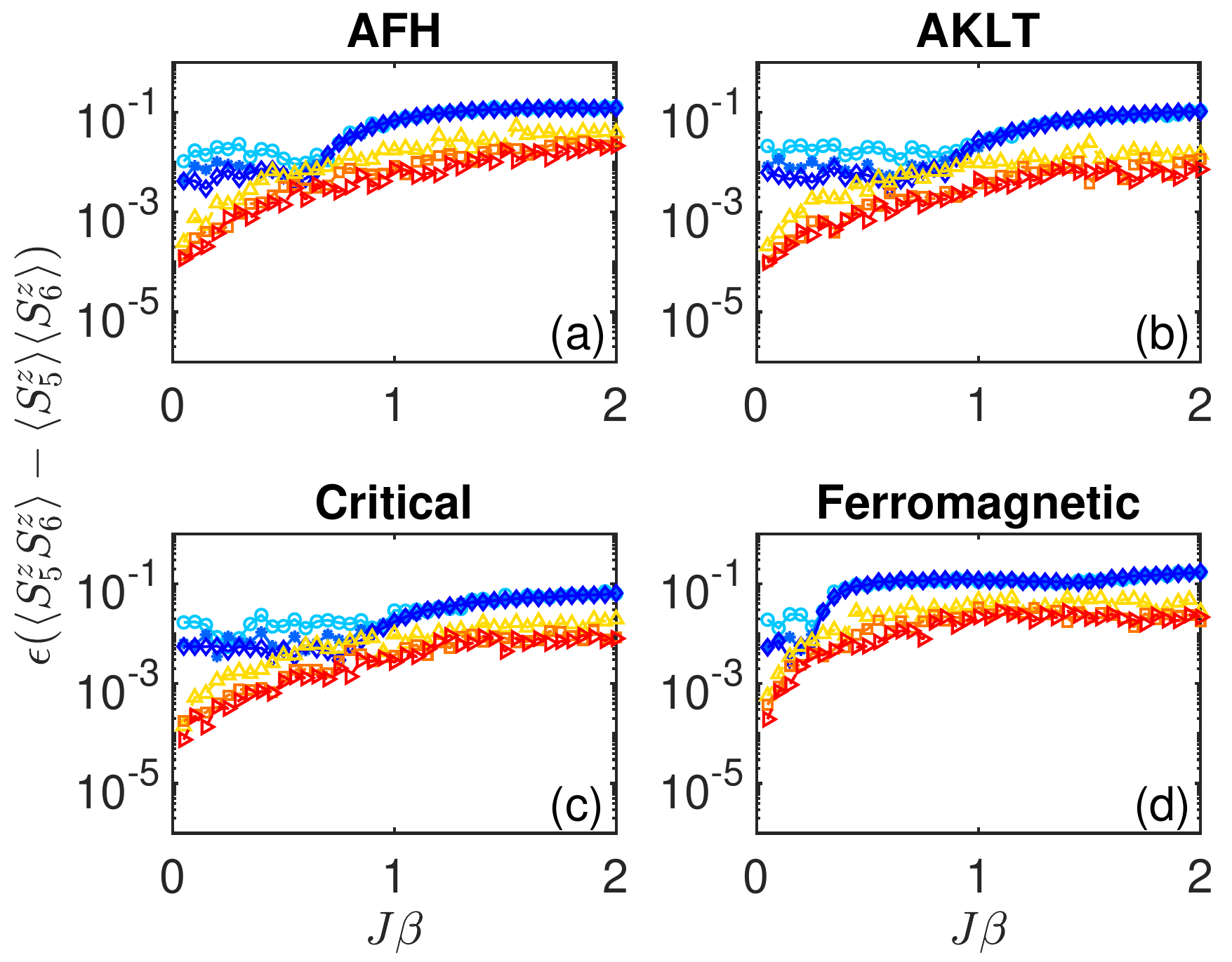}
	\caption{Comparison between CMCS and MCMC of 1D spin-1 bilinear-biquadratic model (\ref{eq:blbq}) with 12 spins for the local region size $\ell = 4$ with sites 4-6, and with different temperatures. Results are shown for varying sample sizes $N_s$ and numbers of unique configurations $N_u$. The panels show the absolute error $\epsilon$ of the correlation between the two central sites $\langle S^z_5 S^z_6\rangle - \langle S^z_5 \rangle\langle S^z_6 \rangle$. The system is in the (a) antiferromagnetic-Heisenberg(AFH) phase when $\theta=0$; (b) Affleck-Kennedy-Lieb-Tasaki(AKLT) phase when $\theta=\arctan(1/3)$; (c) critical phase when $\theta=\pi/4$; and (d) ferromagnetic phase when $\theta=2\pi/3$.
    Blue markers indicate the different samples in MC: 
    \textcolor{blue1}{$\circ$} represents $N_s = 1000$; 
    \textcolor{blue2}{$\times$} represents $N_s = 5000$; 
    \textcolor{blue3}{$\lozenge$} represents $N_s = 10000$. 
    Red markers indicate the different samples in CMCS: 
    \textcolor{red1}{$\triangle$} represents $N_s = 10$ with $N_u \le 810$; 
    \textcolor{red2}{$\square$} represents $N_s = 50$ with $N_u \le 4050$; 
    \textcolor{red3}{$\triangleright$} represents $N_s = 100$ with $N_u \le 8100$. 
    } 
 \label{fig5}
\end{figure}

\section{Thermal States} \label{sec:thermalstate}

In this section, we then investigate the performance of CMCS in thermal states. We consider a spin-1 bilinear-biquadratic chain with Hamiltonian
\begin{align}
    \mathcal{H} = \sum_{i} J\left[ \pmb{S}_i\cdot\pmb{S}_{i+1} + \tan(\theta)(\pmb{S}_i\cdot\pmb{S}_{i+1})^2 \right]\label{eq:blbq}
\end{align}
Where $\theta$ is an angular variables, and $\pmb{S}_i = (S^x_i, S^y_i, S^z_i)$ is the vector of spin-1 operators at the site $i$ \cite{parkinson1988s, binder2020low}. The model can be tuned by varying the angle $\theta$: when $-\pi/4<\theta<\pi/4$, the system is in the Haldane phase; when $\pi/4<\theta<\pi/2$, the system is in the critical phase; when $\pi/2\le\theta<5\pi/4$, the system is in the ferromagnetic phase, and when $\theta = \pi/4, \pi/2$, the system is in the phase transition point. When $\theta=0$, the Haldane characterizes the antiferromagnetic-Heisenberg (AFH) state, and when $\theta=\arctan(1/3)$, the system is described by a topological order, which is the Affleck-KennedyLieb-Tasaki (AKLT) state \cite{affleck1987rigorous, affleck1988valence}. 
The AKLT state is a valence-bond solid on a spin-1 chain. At each site, the spin-1 is decomposed into two spin-1/2s, which are projected onto the symmetric ($S=1$) triplet subspace; on every nearest-neighbor bond, one spin-1/2 from each site forms a singlet. The AKLT state is an exact MPS with bond dimension $D=2$, in which connected two-point correlators decay exponentially with distance. For open chains, the unpaired edge spin-1/2s give rise to characteristic edge states.

The state of a quantum system in thermal equilibrium is described by the density matrix
\begin{align}
    \rho_{\beta} = \frac{e^{-\beta H}}{Z_{\beta}}
\end{align}
where $\beta$ is the inverse temperature, $Z_{\beta}\equiv \text{Tr}(e^{-\beta H})$ is the partition function, and we consider all the symmetry sectors. The expectation value of an observable $O$ is 
\begin{align}
    \langle O \rangle = \text{Tr}(\rho_{\beta}O). 
\end{align}
In the high-temperature limit ($\beta\rightarrow0$),
\begin{align}
    \rho_{\beta}\rightarrow \frac{\mathbb{I}}{\mathcal{N}}
\end{align}
where $\mathcal{N = \dim H}$, i.e, the system approaches the uncorrelated state $\propto \mathbb{I}$ with very small correlations. In the low-temperature limit ($\beta\rightarrow\infty$), 
\begin{align}
    \rho_{\beta}\rightarrow|\psi_{gs}\rangle\langle\psi_{gs}|
\end{align}
the system is increasingly dominated by the ground state $|\psi_{gs}\rangle$.
For the diagonal local observables $\langle S^z_i \rangle$ and $\langle S^z_iS^z_{j} \rangle$, we work in $S^z$ basis $\{ \pmb{m} \}$, where $|\pmb{m}\rangle = |m_1, m_2, ..., m_L\rangle$ with $m_i\in\{+1,0,-1\}$. We then use the probabilities $P_{\beta}(\pmb{m}) = \langle\pmb{m}|e^{-\beta \mathcal{H}}| \pmb{m}\rangle$ for evaluating these observables.

In Fig.~\ref{fig5}, we compare the performance of CMCS and MCMC on the 1D bilinear-biquadratic chain (\ref{eq:blbq}), showing the mean absolute error of the correlator $\langle S^z_5S^z_{6}\rangle $ as a function of $\beta$ in four regimes(AFH, AKLT, critical, and ferromagnetic). 
For all the values of $\theta$ considered, the correlator considered is not trivially $0$. 
We choose a local region of size $\ell=4$ which includes the sites 4-7, 
while the remaining 8 spins form the environment. 
We can see that for MCMC, increasing the number of samples from $N_s = 10^3$ to $N_s = 10^4$ only leads to a modest improvement in the error. In contrast, the CMCS error visibly decreases as we increase $N_s$ from 10 to 100 (hence increasing the number of unique configurations $N_u$ in ranges from $810$ to $8100$ depending on the value of $\theta$). In other words, for comparable effective sample size $N_u \sim N_s$, CMCS achieves much smaller errors across most temperatures.

\section{Conclusion} \label{sec:conclusions} 
In this work, we introduced the CMCS algorithm for the estimation of local observables in quantum spin chains, and we showed that it can allow, in some regimes, to obtain more accurate results compared to estimation from conventional sampling. 
We first tested CMCS on the ground state of a spin-1/2 chain and evaluated its performance using local observables around the central sites in different phases. In both ordered phases, such as the AFM and FM, CMCS achieves a substantial reduction of absolute error compared with conventional Monte Carlo methods, corresponding to a significantly larger effective sample size. Near critical regions, where the correlation length is larger, the advantage of CMCS is reduced or even lost, reflecting the enhanced sensitivity of local observables to distant sites.
We further applied CMCS to thermal states of the spin-1 chain and found that CMCS can outperform standard MCMC for local observables, especially at higher temperature.  

We comment that CMCS does not require significant overhead to be used. The overhead comes from evaluating the wave function for all the unique samples, which will thus require $(2^\ell-1)N_e$ more evaluation from the ones to obtain the $N_s$ samples from MCMC, something that in our case is negligible, and also something that can be readily parallelized.  
As for the evaluation of $\tilde{P}(\pmb{u})$ for the unique sample $\pmb{u}$, it can be executed readily by storing the values of $P(\pmb{u})$.   

In future, we plan to use CMCS for evaluating local observables and gradients in neural network quantum states computations. This could potentially lead to a speed up of the algorithm both for the estimation of expectation values, and for the convergence to optimal wave functions, at least in regimes away from criticality. Furthermore, the concentrated sampling does not have to require that all configurations in the local region are considered, but instead that a much larger portion of all the configurations in the local region is accounted for, compared to the configurations in the environment. This could allow to extend the size of the local region, and/or reduce the total number of configurations. 
Last, here we considered ground states of a Hamiltonian which is not number-conserving, and thermal states for an Hamiltonian with number conservation and contributions from all sectors. In future works, we will extend concentrated sampling to number conserving systems.

\section{Acknowledgements} 
We acknowledge fruitful discussions with E. Khatami, E. Merali and R.T. Scalettar. We also acknowledge the support of the Ministry of Education, Singapore, under the grant MOE-T2EP50123-0017, and from the Centre for Quantum Technologies grant CQT$\_$SUTD$\_$2025$\_$01. 
Part of the numerical work was performed at the National Supercomputing Centre, Singapore \cite{nscc}.

\normalem
\bibliography{Bibliography.bib}

\makeatletter
\close@column@grid

\clearpage

\appendix
\onecolumngrid 

\section*{Appendix}
\section{Estimation of generic operators}\label{app}

Here, we estimate the expectation value of a generic operator $O$ with respect to wave function $|\psi\rangle=\sum_{\pmb{x}} \psi(\pmb{x})\,|\pmb{x}\rangle$ expressed in the
computational basis,. It is given by

\begin{equation}
	\begin{aligned}
		\langle O \rangle =& \frac{\langle \psi|O|\psi\rangle}{\langle \psi|\psi \rangle}\\
		=&  \sum_{\pmb{x}} \frac{\psi^*(\pmb{x})}{\langle\psi|\psi\rangle} \sum_{\pmb{x}'}O_{\pmb{x},\pmb{x}'}\psi(\pmb{x}')\\     
		=& \sum_{\pmb{x}} \frac{|\psi^*(\pmb{x})|^2}{\langle\psi|\psi\rangle} \sum_{\pmb{x}'}O_{\pmb{x},\pmb{x}'}\frac{\psi(\pmb{x'})}{\psi(\pmb{x})}\\  
		=& \sum_{\pmb{x}}P(\pmb{x}) O_{loc}(\pmb{x})
	\end{aligned}
\end{equation}
In the CMCS scheme, we renormalize the wave function on the set $\mathcal U$ of unique configurations generated by the chosen local region. It reads
\begin{equation}
	\begin{aligned}
		\widetilde{\psi}(\pmb{u}) = \frac{\psi(\pmb{u})}{\sqrt{\sum_{\pmb{u}} \psi^*(\pmb{u})\psi(\pmb{u})}}
	\end{aligned}
\end{equation}
and the renormalized probability is 
\begin{equation}
	\begin{aligned}
		\widetilde{P}(\pmb{u}) = \frac{|\psi(\pmb{u})|^2}{\sum_{\pmb{u}} \psi^*(\pmb{u})\psi(\pmb{u})}
	\end{aligned}
\end{equation}

Thus, the expectation value of the operator in CMCS method is 
\begin{equation}
	\begin{aligned}
		\langle O \rangle =& \sum_{\pmb{u} \in \mathcal{U}}\widetilde{P}(\pmb{u}) \sum_{\pmb{u}'}O_{\pmb{u},\pmb{u}'}\frac{\psi(\pmb{u'})}{\psi(\pmb{u})}
	\end{aligned}
\end{equation}

\begin{figure}
	\centering
	\includegraphics[width=0.7\linewidth]{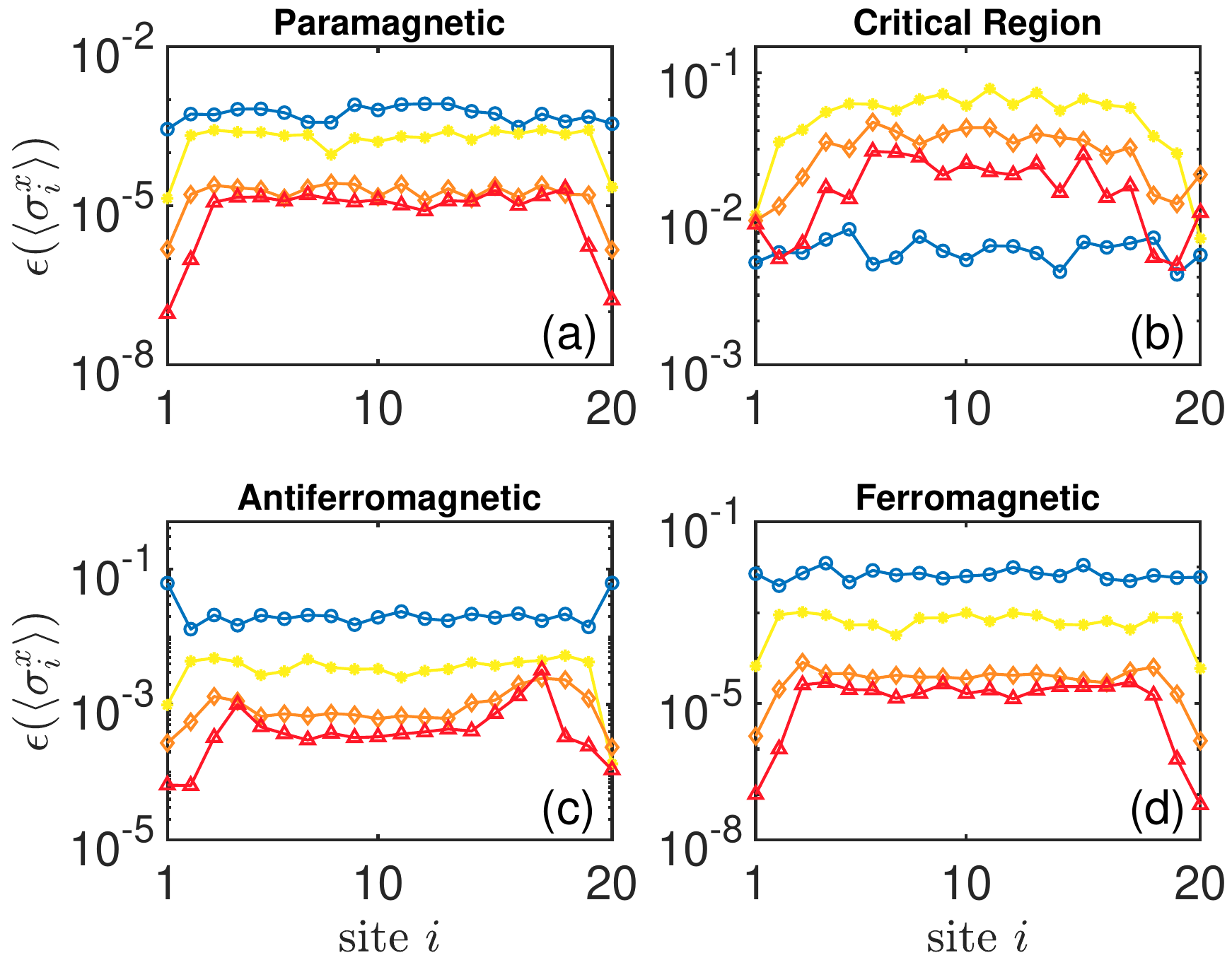}
	\caption{Comparison of the mean absolute error of operator 
		$\langle \sigma^x_i \rangle$ between MCMC and CMCS for the ground state of 1D 20 spins tilted Ising model using different local-region sizes $\ell$. The system is in 
		(a) the paramagnetic phase with $h_x = 10J, h_z = J/2$, 
		(b) the critical region with $h_x=0.95J, h_z=J/2$, 
		(c) the antiferromagnetic phase with $h_x=J/2, h_z = J/2$, and 
		(d) the ferromagnetic phase with $J<0, h_x = |J|/2, h_z = |J|/2$. The local region is shifted along the chain as the site $i$ is varied. For each local region, the measurement site where $\langle \sigma_i^x\rangle$ is measured is chosen as follows: for $\ell=2$, we use the first site of the region; for $\ell=3$, we use the central site of the region; and for $\ell=4$, we use the two central sites of the region (measurements at the boundary sites of the local region are also considered).
		\textcolor{fig2blue}{$\circ$} represents MC sampling with $N_s=10^4$;
		\textcolor{fig2red1}{$\ast$} represents CMCS $\ell = 2$ ;
		\textcolor{fig2red2}{$\lozenge$} represents CMCS $\ell = 3$ ;
		\textcolor{fig2red3}{$\bigtriangleup$} represents CMCS $\ell = 4$.
	} 
	\label{fig6}
\end{figure}

For diagonal operators ($O_{\pmb{x,x'}}=O(\pmb{x})\delta_{\pmb{x,x'}}$), only the configuration $\pmb{x'}=\pmb{x}$ contributes, and no extra configurations need to be included; off-diagonal operators, instead, require that all connected states $\pmb{x}'$ are contained in $\mathcal{U}$. Fig. \ref{fig6} shows the mean absolute error of $\langle \sigma_i^x\rangle$ as an example off-diagonal observable.
Since site $i$ must lie in the local region, the local region is shifted along the chain as $i$ is varied. We observe that its performance is comparable to that of the diagonal observable $\langle\sigma_i^z\rangle$, depicted in Fig. \ref{fig2}.

\end{document}